\documentclass[twocolumn,showpacs,showkeys]{revtex4}
\usepackage{graphicx}
\usepackage{bm}
\usepackage{color}
\usepackage{amsmath}
\usepackage{natbib}

\begin{document}

\title{Surface spin-electron acoustic waves in magnetically ordered metals}

\author{Pavel A. Andreev}
\email{andreevpa@physics.msu.ru}
\author{L. S. Kuz'menkov}%
\email{lsk@phys.msu.ru}
\affiliation{Faculty of physics, Lomonosov Moscow State University, Moscow, Russian Federation.}

 \date{\today}

\begin{abstract}
Degenerate plasmas with motionless ions show existence of three surface waves: the Langmuir wave, the electromagnetic wave, and the zeroth sound. Applying the separated spin evolution quantum hydrodynamics to half-space plasma we demonstrate the existence of the surface spin-electron acoustic wave (SSEAW). We study dispersion of the SSEAW. We show that there is hybridization between the surface Langmuir wave and the SSEAW at rather small spin polarization. In the hybridization area the dispersion branches are located close to each other. In this area there is a strong interaction between these waves leading to the energy exchange. Consequently, generating the Langmuir waves with the frequencies close to hybridization area we can generate the SSEAWs. Thus, we report a method of creation of the SEAWs.
\end{abstract}


\maketitle


Plasmonics is a great field of technological application of plasmas \cite{Barnes Nat 03}, \cite{Giannini CR 11}. Key role in plasmonics belongs to the surface Langmuir waves or surface plasmons \cite{Fang APL 08}, \cite{Wang APL 11}. On the over hand, the materials with the partial spin polarization of the carriers reveal the existence of spin-electron acoustic waves (SEAWs) \cite{Andreev PRE 15}, \cite{Andreev AoP 15}. Quantum of the SEAWs is called spelnon. The spelnons in the bulk materials exist simultaneously with the bulk plasmons, but the spelnons have smaller energies than the energy of plasmons as it is shown in Ref. \cite{Andreev PRE 15}. SEAWs in 2D structures, they are also called spin-plasmons, are considered in Refs. \cite{Andreev spin-up and spin-down 1408 2D}-\cite{Agarwal PRB 14}. Their spectrum for plane-line two-dimensional structures is described in Refs. \cite{Andreev spin-up and spin-down 1408 2D}, \cite{Ryan PRB 91} for the external magnetic field perpendicular to the plane. Contribution of the cyclotron motion in the spectrum is described in \cite{Andreev spin-up and spin-down 1408 2D}. Influence of the disorder on properties of spin-plasmons is described in Ref. \cite{Agarwal PRB 14}. A possibility of the spin-electron acoustic soliton formation due to the non-linear evolution of perturbations in the partially spin polarized electron gas was demonstrated in Ref. \cite{Andreev 1504}. It was shown that the electron-spelnon interaction leads to the  Cooper pair formation ensuring a mechanism for the high-temperature superconductivity \cite{Andreev HTSC 15}. Experimental analysis of surface spin waves and spin waves in thin films is a subject of current research (see for instance \cite{Michel PRB 15}) leading to a possibility of the experimental analysis of the SEAWs.

Moving towards applications of the SEAWs, in this paper, we study a possibility of existence of the surface SEAWs (SSEAWs) and their properties. Surface waves in plasmas were studied long time ago \cite{Tend PRL 67}-\cite{Jones PRL 83}. However, the zero-sound mode, which is well-known for the bulk degenerate perturbations of fermions (see for instance \cite{Landau v9} and \cite{JETP 80th}), was just recently described for the half space regime \cite{Tyshetskiy JPP 13}, \cite{Tyshetskiy PP 14}.
Collisional and collisionless Landau damping of the surface Langmuir waves in the degenerate electron gas are considered in \cite{Tyshetskiy PP 12}.
There is still the fundamental interest to the quantum effects caused by the quantum Bohm potential in surface plasma waves \cite{Lazar PP 07}-\cite{Moradi PL A 15}.

\begin{figure}
\includegraphics[width=8cm,angle=0]{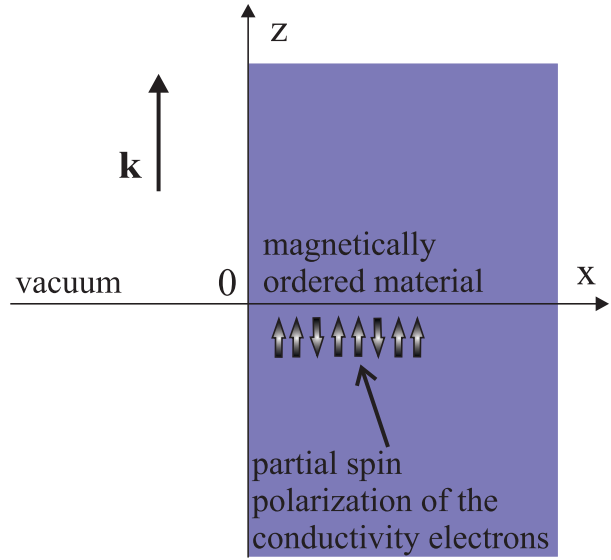}
\caption{\label{SSeaw} (Color online) The figure shows the half spaced magnetically ordered material considered to find the surface SEAWs.}
\end{figure}

To find the SEAW a new method called the separate spin evolution quantum hydrodynamics
(SSE-QHDs) was developed in Ref. \cite{Andreev PRE 15}. This method is a
generalization of usual spin-1/2 QHD developed in Refs. \cite{MaksimovTMP 2001}, \cite{Marklund PRL07}, \cite{Andreev VestnMSU 2007} (some applications are described in review \cite{Shukla RMP 11}). A generalization of the SSE-QHD was developed in Ref. \cite{Trukhanova PLA 15}, where the spin current evolution equation was derived in terms of separate evoltion of spin-up and spin-down electrons.
The SSE-QHD was developed in Ref. \cite{Andreev PRE 15}, however it does not consider the thermal part of the spin current. The thermal part of the spin current was recently derived in Ref. \cite{Andreev TerSpinCurent} for the degenerate electron gas, where the thermal part of the spin current is reduced to the Pauli blocking contribution, similarly to the Fermi pressure.

We do not present full set of the SSE-QHD here, we restrict our presentation with the linearized equations for the wave propagating parallel to the boundary surface. Full set of the SSE-QHD equations can be found in Refs. \cite{Andreev PRE 15}, \cite{Andreev AoP 15}.
To describe the surface SEAWs we need the continuity and Euler equations for the separate spin evolution of the spin-up and spin-down electrons.
Corresponding linearized and Fourier transformed set of equations arises as
$\omega\delta n_{s}=n_{0s}k_{z}\delta v_{sz}$, $-mn_{0s}\omega\delta \textbf{v}_{s}+(6\pi^{2})^{\frac{2}{3}}n_{0s}^{\frac{2}{3}}\hbar^{2}k_{z}\delta n_{s}\textbf{e}_{z}/3m=q_{e}n_{0s}\delta \textbf{E}$ we solve the hydrodynamic equations in each half space independently. We consider them
together with the Maxwell equations
$\omega\delta B_{y}=k_{z}ñ\delta E_{x}+\imath ñ\partial_{x}\delta E_{z}$,
$\omega\delta E_{x}=k_{z}ñ\delta B_{y}+4\pi\imath(n_{0u}\delta v_{ux}+n_{0d}\delta v_{dx})$,
and
$\omega\delta E_{z}=\imath ñ\partial_{x}\delta B_{y}+4\pi\imath(n_{0u}\delta v_{uz}+n_{0d}\delta v_{dz})$ solved in whole space.

We consider the following equations of state for the spin-up and spin-down electrons $P_{s}=(6\pi^{2})^{2/3}n_{0s}^{5/3}\hbar^{2}/5m$. These equations include the deformation of the Fermi step distribution under the action of the external magnetic field or the strong exchange interaction in the magnetically ordered materials.

The effect of spin polarization on the Fermi pressure has been considered in literature in the application to the wave phenomena in the single
fluid model of electrons \cite{JETP 80th}, \cite{Andreev AoP 14} and the two-fluid model \cite{Andreev PRE 15}, \cite{Andreev spin-up and spin-down 1408 2D}, \cite{Ryan PRB 91}.

We consider the regime of zero magnetic field and assume that the partial spin polarization of conducting electrons in magnetic materials is caused by inner equilibrium effects like the exchange interaction.
To describe the main properties of the SSEAWs we consider a surface of the ferromagnetic or ferrimagnetic materials containing the conducting electrons with the partial spin polarization as it is shown in Fig. 1.

Our model works if both subspecies of electrons are degenerate: $T\ll T_{Fu},T_{Fd}$, where $T$ is the temperature of the system and $T_{Fs}=(6\pi^{2}n_{0s})^{2/3}\hbar^{2}/2m$ are the Fermi temperatures for the spin-up and spin-down electrons. It gives a restriction on the large spin polarizations $\eta=|n_{0d}-n_{0u}|/(n_{0d}+n_{0u})$, since $T_{Fu}$ would be very small and system could be described at low temperatures only. At the high spin polarization contributions of the Coulomb exchange interaction between the conductivity electrons and ion dynamics also affect evolution of electrons. The concentration of conductivity electrons in metals is large $n_{0}\sim 10^{22}$ cm$^{-3}$, hence at the spin polarization $\eta$ below $\eta_{0}=0.99$ we can consider temperatures in area from 10 K to 300 K.

Hydrodynamic equations give the following expressions for the velocity field in terms of the electric field perturbation
$\delta v_{sx}=-(e/m)\imath\delta E_{x}/\omega,$
and
$\delta v_{sz}=-(e/m)\imath\omega\delta E_{z}/(\omega^{2}-k_{z}^{2}U_{s}^{2}),$
where $U_{s}^{2}=\frac{\hbar^{2}}{3m_{e}^{2}}(6\pi^{2}n_{0\uparrow})^{\frac{2}{3}}$. These formulae can be substituted to the Maxwell equations to find an equation for electromagnetic field in our system.

At the zero external magnetic field and zero spin polarization the dielectric permittivity arises as a diagonal tensor with the following nonzero elements $\varepsilon_{xx}=\varepsilon_{yy}\neq\varepsilon_{zz}$. If we include the spin polarization the structure of the dielectric tensor is the same, but the explicit form of $\varepsilon_{zz}$ modifies becoming more complex:
\begin{equation}\label{SSeaw} \varepsilon_{ud}\equiv\varepsilon_{zz}=1-\frac{\omega_{Lu}^{2}}{\omega^{2}-k_{z}^{2}U_{u}^{2}}-\frac{\omega_{Ld}^{2}}{\omega^{2}-k_{z}^{2}U_{d}^{2}}, \end{equation}
while
\begin{equation}\label{SSeaw}\varepsilon\equiv\varepsilon_{xx}=\varepsilon_{yy}=1-\frac{\omega_{Le}^{2}}{\omega^{2}} \end{equation}
is not changed. Here $\omega_{Ls}^{2}=4\pi e^{2}n_{0s}/m$ are the partial Langmuir frequencies, with $s=u,d$ for the spin-up and spin-down electrons, and $\omega_{Le}^{2}=4\pi e^{2}n_{0e}/m$ is the full Langmuir frequency, $n_{0}=n_{0u}+n_{0d}$ is the full equilibrium concentration of the electrons. Difference of the equilibrium concentrations arises due to the spin polarization $\eta$, hence $n_{0u}=(1-\eta)n_{0}/2$ and $n_{0d}=(1+\eta)n_{0}/2$.

The surface waves under consideration are described by
\begin{equation}\label{SSeaw equation} c^{2}\partial_{x}^{2}\delta E_{z}+\omega^{2}\frac{\varepsilon_{ud}}{\varepsilon}\biggl(\varepsilon-\frac{k^{2}c^{2}}{\omega^{2}}\biggr)\delta E_{z}=0. \end{equation}
It describes waves of $E$-type.
In vacuum, at $x<0$, we have $\varepsilon_{ud}=\varepsilon=1$

Equation (\ref{SSeaw equation}) gives the following solution for left and right half-spaces:
\begin{equation}\label{SSeaw} \delta E_{z}=\Biggl\{\begin{array}{cc}
                                               C_{1}\exp\biggl(-\sqrt{\frac{\varepsilon_{ud}}{\varepsilon}k_{z}^{2}-\varepsilon_{ud}\frac{\omega^{2}}{c^{2}}}x\biggr), & x>0 \\
                                               C_{2}\exp\biggl(\sqrt{k_{z}^{2}-\frac{\omega^{2}}{c^{2}}}x\biggr), & x<0.
                                             \end{array}
 \end{equation}

Boundary conditions are the continuity of $\delta E_{z}$ and $\delta B_{y}$
\begin{equation}\label{SSeaw} \begin{array}{cc}
\{\delta E_{z}\}\mid_{x=0}=0, & \{\delta B_{y}\}\mid_{x=0}=0.
\end{array} \end{equation}

The linearized Maxwell equations presented above give the following relation between $\delta B_{y}$ and $\delta E_{z}$:
\begin{equation}\label{SSeaw By} \delta B_{y}=\frac{\imath c}{\omega} \frac{\varepsilon}{\varepsilon-\frac{k^{2}c^{2}}{\omega^{2}}}\partial_{x}\delta E_{z}.\end{equation}

Condition $\{\delta E_{z}\}\mid_{x=0}=0$ leads us to $C_{1}=C_{2}$. The following application of the condition $\{\delta B_{y}\}\mid_{x=0}=0$ gives the following dispersion equation:
\begin{equation}\label{SSeaw disp eq polarized} \sqrt{k_{z}^{2}c^{2}-\omega^{2}}+\varepsilon\sqrt{\frac{\varepsilon_{ud}}{\varepsilon}}\sqrt{k_{z}^{2}c^{2}-\varepsilon\omega^{2}}=0. \end{equation}
This equation has real solutions if $\varepsilon<0$ and $\varepsilon_{ud}<0$.

Dispersion equation (\ref{SSeaw disp eq polarized}) arises at the application of the full set of the Maxwell equations. Coupling of waves appears in the limit of the large wave vectors $k$ as we show it below at the numerical analysis. Hence, it corresponds to the quasi-electrostatic regime. We can simplify the dispersion equation (\ref{SSeaw disp eq polarized}) in this regime to $1-\varepsilon\varepsilon_{ud}=0$.

Equation $1-\varepsilon\varepsilon_{ud}=0$ is a generalization of the dispersion equation existing for unpolarized degenerate electron gas, in the quasi-electrostatic regime,
\begin{equation}\label{SSeaw disp eq ElStat with pr no polariz} 1-\biggl(1-\frac{\omega_{Le}^{2}}{\omega^{2}}\biggr)\biggl(1-\frac{\omega_{Le}^{2}}{\omega^{2}-\frac{1}{3}v_{Fe}^{2}k^{2}}\biggr)=0, \end{equation}
which has the following solution $\omega^{2}=0.5(\omega_{Le}^{2}+v_{Fe}^{2}k^{2}/3)$.

Usually we have two branches of dispersion dependence of the surface waves, since the hydrodynamics does not describe the zero sound. One of them is the Langmuir wave, which is highly nonlongitudinal wave in contrast with its bulk analog. Its frequency tends to $kc$ at $k\rightarrow0$. In the opposite limit of large wave vectors $k\rightarrow\infty$ (its frequency tends to $\omega_{Le}/\sqrt{2}$ if we drop the pressure contribution) this wave become longitudinal and can be described in the quasistatic limit. The second branch is the analog of the three dimensional electromagnetic wave.

Separate spin evolution leading to equation (\ref{SSeaw disp eq polarized}) gives three branches of the dispersion dependence. New branch is related to the spin polarization. It is the surface SEAWs or surface spelnons, similar to the bulk SEAWs considered in Refs. \cite{Andreev PRE 15}, \cite{Andreev AoP 15}.

Fig. 2 shows that at rather large spin polarization the SEAW dispersion dependence lies below the Langmuir wave. Decrease of the spin polarization increases the frequency of the SEAW. Thus, we find crossing of the dispersion dependencies of the Langmuir wave and the SEAW, which leads to the hybridization of their spectrums as it shown in Fig. 3. In figures we use the dimensionless frequency $\Omega\equiv\omega/\omega_{Le}$ and the dimensionless wave vector $\kappa=v_{Fe}k_{z}/(\omega_{Le}\sqrt{3})$, where $v_{Fe}=(3\pi^{2}n_{0})^{\frac{1}{3}}\hbar/m$ is the Fermi velocity.

\begin{figure}
\includegraphics[width=8cm,angle=0]{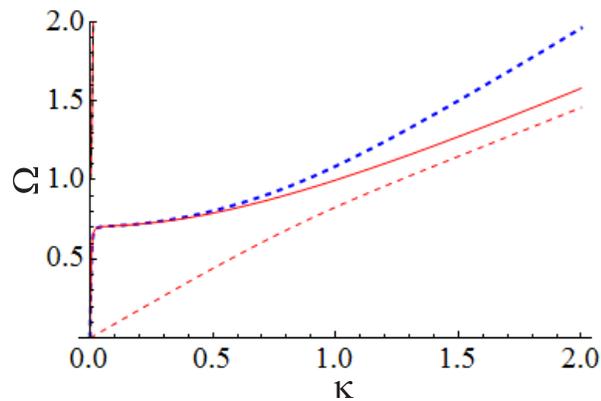}
\caption{\label{SSeaw_01_usual} (Color online) This figure describes the spectrum of the surface waves in the partially
spin polarized electron gas. The three dashed lines show the dispersion dependencies.
The slightly visible branch located close to the vertical axis describes the
electromagnetic surface wave in the half-space plasmas. Its frequency grows very
fast at the increase of the wave vector relatively to other waves and it is not
affected by the effects considered in this paper. The continuous line presents the
surface Langmuir wave with no account of the spin polarization in the Fermi pressure.
The blue dashed line above it describes the Langmuir wave with the account of the
spin polarization. The lowest dashed line describes the SSEAW. At the small wave
vectors the Langmuir wave have well-known linear spectrum $\omega\approx kc$.}
\end{figure}

\begin{widetext}
Analytical solutions for the Langmuir wave and the SEAW can be found in the quasi-electrostatic limit
$$\omega^{2}=\frac{1}{4}\Biggl\{\omega_{Le}^{2}+k_{z}^{2}\biggl[U_{u}^{2}\biggl(\frac{n_{0}+n_{0d}}{n_{0}}\biggr)+U_{d}^{2}\biggl(\frac{n_{0}+n_{0u}}{n_{0}}\biggr)\biggr]$$
\begin{equation}\label{SSeaw An Sol ESLim} \pm\sqrt{\Biggl[\omega_{Le}^{2}+k_{z}^{2}\biggl[U_{u}^{2}\biggl(\frac{n_{0}+n_{0d}}{n_{0}}\biggr)+U_{d}^{2}\biggl(\frac{n_{0}+n_{0u}}{n_{0}}\biggr)\biggr]\Biggr] -8k_{z}^{4}U_{u}^{2}U_{d}^{2}-8k_{z}^{2}(\omega_{Lu}^{2}U_{d}^{2}+\omega_{Ld}^{2}U_{u}^{2})}\Biggr\}. \end{equation}
\end{widetext}

\begin{figure}
\includegraphics[width=8cm,angle=0]{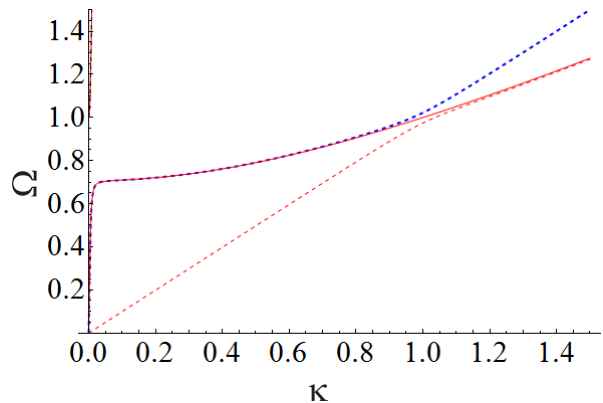}
\caption{\label{SSeaw_01_usual} (Color online) The figure shows the spectrum of the surface waves, but for smaller spin polarizations, in compare with Fig. 2. Here we present a regime of hybridization of spectrum of the Langmuir and spin-electron acoustic waves.}
\end{figure}

At the small wave vectors we numerically find the linear dispersion dependence for the surface SEAWs. Therefore, we find an approximate analytical solution of equation (\ref{SSeaw disp eq polarized}) for this regime
\begin{equation}\label{SSeaw An sol appr small k} \Omega^{2}=\frac{1}{2}[(1-\eta)(1+\eta)^{\frac{2}{3}}+(1+\eta)(1-\eta)^{\frac{2}{3}}]\xi^{2}, \end{equation}
which coincides with the dispersion dependence of the bulk SEAWs in the small wave vector regime.

Numerical analysis shows that the Fermi pressure and the spin separation do not affect the high frequency electromagnetic wave. The separate spin evolution does not affect it either. On the other hand spin separation could affect other branches. For instance the surface Langmuir waves is affected by the modification of the Fermi pressure at the spin polarization.

The spin polarization increases the Fermi pressure contributing as $P_{pol}=P_{u}+P_{d}=0.5[(1+\eta)^{5/3}+(1-\eta)^{5/3}]P_{Fe}$ (see \cite{Andreev PRE 15}, \cite{JETP 80th}, \cite{Andreev AoP 14}), where $P_{Fe}=(3\pi^{2})^{2/3}n_{0}^{5/3}\hbar^{2}/5m$ is the Fermi pressure at the zero spin polarization. At the full spin polarization $P_{pol}$ increases up to $1.6\cdot P_{Fe}$. Therefore, it increases the frequency of the surface Langmuir wave.

We can explicitly see this effect in Fig. 2. It reveals in the shift of the dashed blue line from the continuous red line.

For small spin polarization this effects rather small. Consequently, we see coincidence of the continuous red line and dashed red line in Fig. 3 at the large wave vectors $k$.

The area of the spectrum hybridization is located at the wave vectors $\kappa\approx\kappa_{0}=1$. Hybridization reveals in the fact that the Langmuir wave branch at $\kappa\ll\kappa_{0}$ fuses with the SEAW wave branch at $\kappa\gg\kappa_{0}$ and the SEAW branch at $\kappa\gg\kappa_{0}$ continuously turn in the Langmuir wave branch at $\kappa\ll\kappa_{0}$. Switching of branches happens at $\kappa\approx\kappa_{0}$. Branches of the Langmuir waves and the SEAWs would cross each other, but interaction of waves changes the structure of the spectrum.

Area of hybridization of the spectrum of two waves shows an area of strong interaction of these waves. In this area two waves are similar to two bound pendulums. Excitation of one of waves with $\kappa\approx\kappa_{0}$ leads to the energy transition to the second wave. Thus, excitation of the Langmuir waves and the increase of their energy to reach point $(k_{0}, \omega(k_{0}))$ allows to generate the SSEAWs. This is the first report on the possibility of the SEAW generation, while several papers have been published to describe properties of two- and three-dimensional SEAWs \cite{Andreev PRE 15}, \cite{Andreev AoP 15}, \cite{Ryan PRB 91}, \cite{Agarwal PRL 11}.

The SEAWs are longitudinal waves, where concentration of the spin-up and spin-down electrons oscillate with opposite phase. It leads to the local collective motion of the spin-up and spin-down fluids in opposite directions and may be considered as a spin wave in the electron gas. Mostly, consideration of spin waves are related to the magnetic moments of lattice containing bound electron on unfilled d and f shells, even at the analysis of conducting materials. The SEAWs are examples of spin waves in the polarized electron gas along with other transverse spin waves reported in Refs. \cite{Andreev VestnMSU 2007}, \cite{Andreev TerSpinCurent}.

Surface spelnons simultaneously show similarity and difference to the plasmons. Both of them have linear spectrum at the small wave vectors. However, plasmons dynamics is related to the electrodynamic effects and their phase velocity is close to the speed of light. Surface plasmons strongly interact with the small frequency light $\omega<\omega_{Le}$, which cannot penetrate in metals. Spelnons are also related to the electron oscillations, but the spelnons are longitudinal even at $k\rightarrow0$. Judging on this property they show similarity to phonons (see an example of resent work on phonons \cite{Schuetz PRX 15}), but the spelnons have intermediate phase velocities $c_{sn}\sim0.01 c$. Having spin nature and having intermediate properties between phonons and surface plasmons the spelnons become potentially useful and interesting subject of research.

Hybridization of the dispersion branches occurring at the linear wave coupling is a widely spread phenomenon \cite{Zheleznyakov SPU 83}. It happens in many cases including plasma physics \cite{Bogdankevich SUP 81}, \cite{Birau PU 97}, appearance of exciton-polaritons in semiconductors \cite{Deng RMP 10}, \cite{Byrnes NP 14}, metamaterials \cite{Rupin SR 15}. As we have shown above the SSEAWs in the spin polarized electron gas of ferromagnetic materials reveals a hybridization of the spectrum either.

In conclusion we want summarize that we described spectrum of collective excitations in the half spaced partially spin polarized electron gas of magnetically ordered materials containing the spin-electron acoustic wave and found regime of hybridization of the Langmuir wave and SEAW. We stress attention that strong interaction of waves in area of hybridization of wave dispersion dependencies leads to the energy transition from the Langmuir wave to the SEAW. It gave a mechanism of generation of the SSEAWs.

Work of P.A. is supported by the Dynasty foundation.

\end{document}